# On the stability of a space vehicle riding on an intense laser beam

Popova H.[1,*], Efendiev M.[2], Gabitov I.[3]

[1]National Academy of Aviation, Baku, Azerbaijan, on leave from Skobeltsyn Institute of Nuclear Physics, Lomonosov Moscow State University, GSP-1, Leninskie Gory, Moscow, 119991, Russian Federation;

[2]Helmholtz Zentrum München Deutsches Forschungszentrum für Gesundheit und Umwelt (GmbH) Institute of Computational Biology Ingolstädter Landstr. 1 85764 Neuherberg, Germany;

[3]Skoltech Center for Photonics and Quantum Materials, Building 3, Moscow, 143026, Russian Federation.

[*]Contact: popovaelp@mail.ru

The Breakthrough Starshot Initiative is suggested to develop the concept of propelling a nano-scale spacecraft by the radiation pressure of an intense laser beam. If such a nanocraft could be accelerated to 20 percent of light speed, it could reach the vicinity of our nearest potentially habitable exoplanet within our life time and capture its images and obtain other scientific data. In the case of Proxima Centauri B, a recently discovered habitable zone in the orbit around Proxima Centauri (nearest star system), such a flight would take just 20 years. In this project the nanocraft is a gram-scale robotic spacecraft comprising two main parts: StarChip and Lightsail. The StarChip is a gram-scale wafer, carrying cameras, photon thrusters, power supply, navigation and communication equipment, and constituting a fully functional space probe. The Lightsail is made of extremely thin (no more than a few hundred atoms thick) and light-weight (gram-scale mass) dielectric metamaterial.

To achieve the goal of the project it is necessary to solve a number of outstanding scientific and engineering problems. One of these tasks is to make sure that the nanocraft position and orientation inside the intense laser beam column is stable. The nanocraft driven by intense laser beam pressure acting on its Lightsail is sensitive to the torques and lateral forces reacting on the surface of the sail. These forces influence the orientation and lateral displacement of the spacecraft, thus affecting its dynamics. If unstable the nanocraft might be even expelled from the area of laser beam. In choosing the models for nanocraft stability studies we are using several assumptions: 1. configuration of nanocraft is treated as rigid body (applicability of Euler equations); 2. flat or concave shape of circular sail; 3. mirror reflection of laser beam from surface of the Lightsail. We found conditions of stability for spherical and conical shape of the sail. The simplest stable configurations require the StarChip to be removed from the sail to make the distance to the center of mass of nanocraft bigger than curvature radius of the sail. Stability criteria do not require the spinning of the nanocraft. A flat sail is never stable (even with spinning).

I. Introduction

The conceptual idea to accelerate macroscopic objects by radiation pressure of intense electromagnetic waves has been known since the middle of the 20th century [1-5]. More detailed and up to date arguments [6] have been used to promote interstellar flights using high power lasers. Now such an approach is suggested for solving key problem of nanocraft acceleration in the project "Breakthrough Starshot" (www.breakthroughinitiative.com). The Breakthrough Starshot program is aiming to demonstrate proof of concept for light-propelled nanocrafts capable to fly at 20 percent of light speed and capture images of possible planets and other scientific data in our nearest star system, Alpha Centauri, just over 20 years after their launch. In this project the nanocraft are a gram-scale robotic spacecraft comprising of two main parts: StarChip and Lightsail. The StarChip is a gram-scale wafer, carrying cameras, photon thrusters, power supply, navigation and communication equipment, and constituting a fully functional space probe. The Lightsail is made of extremely thin (no more than a few hundred atoms thick) and light-weight (gram-scale mass) dielectric metamaterial.

The nanocraft driven by intense laser beam pressure acting on its Lightsail is sensitive to the torques and lateral forces reacting on the surface of the sail. This forces influences the orientation and lateral displacement of the spacecraft, thus affecting its dynamics. If unstable the nanocraft might even be expelled from the area of laser beam. The most dangerous perturbations in the position of nanocraft inside the beam and its orientation relative to the beam axis are those with direct coupling between rotation and displacement ("spin-orbit coupling").

Hence, the goal of the study is to determine the conditions at which those disturbances remain tolerable. Here, in considering a problem of such stability of nanocraft illuminated by intense laser beam, we are looking for the simplest possible (but still general enough) models of spacecraft mechanical configurations and the shapes of EM beam. We assume that the nanocraft is a rigid body consisting of Lightsail, StarChip and suspension lines (Fig.1). Rigidity is supported by tensions in suspension lines (as in the case of parachute). This implies that we can use the Euler equations for angular changes (orientation) and equations of motion of the center of mass for lateral displacements. That would require calculation of torques and forces caused by the laser beam. By moving the StarChip further away from the sail we can adjust the moments of inertia of spacecraft configuration and most importantly (as we will show below) significantly change the torques. Most previous attempts to analyze the stability dealt with with the sail being singled out, while mass of the payload was hidden in the sail [7-10]. In this case a stable acceleration by simple shaped EM beam profile would require the spinning of the sail with sufficiently high angular velocity. The only way to achieve stability without spinning the sail would require a more complicated shape of the beam (such as the superposition of 4 Gaussian shaped beams suggested in [11]).

We considered various shapes of sail: a) spherical (coincides with parabolic for small sizes) as most appropriate for final configuration of nanocraft en route; b) conical; c) flat (simplest) (will be seen to be unstable so that even spinning of craft does not help). In our models the laser beam profile is chosen as constant along Z-axis (the direction of propagation), axial symmetry in crossection (in a plane perpendicular Z-axis) with two choices of intensity dependence in that plane : 1) uniform with sharp boundary (when intensity of the beam is $w(\rho)$=const=$w_0$ inside an interval $\rho<a$ and 0 at $\rho>a$, where $\rho$ is the radial distance from the axis of beam; 2) Gaussian (to emulate the diffraction controlled shape of beam). We assume mirror reflection of the laser beam from the surface of the Lightsail and virtually negligible absorption.

In Section II we will analyze stability of nanocraft orientation for different conditions (different Lightsail shape and beam profile). In Section III we will discuss our results.

## II. STABILITY ANALYSIS

### 1. Spherical shape of sail.

The assumptions in choosing the model are: 1) configuration of nanocraft with mass *m* is treated as a solid body (applicability of Euler equations); 2) spherical shape of sail; 3) laser beam intensity is constant inside a cylinder with sharp boundary; 4) mirror reflection of laser beam from surface of the Lightsail.

Scheme of such nanocraft is presented in Fig. 2.

The dynamics of the nanocraft is described by the equations of motion of its center of mass coupled to Euler's equations for rotation around the principal axes. Fig. 3. shows the geometrical arrangement and choices of variables. Rotation around Z axis is not accompanied with the torque associated with angular motions $\theta_x$ and $\theta_y$ and displacements along X and Y (in the initial plane of sail). Hence we are dropping the corresponding component of Euler equation. Instead we might include the presence of fixed angular velocity around Z as a potential stabilizing factor (by spinning).

The nanocraft is considered as a rigid body consisting of the Lightsail connected to the StarChip. The Lightsail is assumed to be a part of spherical surface (of curvature radius *R*) with the mass $m_{ls}$ and *a* as radial extention of sail along the spherical surface. The StarChip is considered as a material point of mass $m_{sc}$ positioned along the primary axis of symmetry of the Lightsail (which is perpendicular to its plane) at the distance of L behind it. Then the center of mass of the nanocraft is at the distance $L_c$ from the center of Lightsail. In the particular case of equal masses ($m_{ls}=m_{sc}$) $L_c=L/2$.

We associate the center of mass with the origin 0 of a Cartesian coordinate system of points **X** = (X; Y; Z)$^T$ such that the Z-axis is directed along the primary axis of cylindrical symmetry of the lightsail. Here $^T$ stands for transpose. In such a way, the lightsail is in the plane which is parallel to the XY-plane. The choice of the center of mass at rest Z=0 signifies that we are dealing with non-inertial frame of reference moving along the axis of laser beam Z with acceleration depending on ponderomotive force of beam $m\ddot{Z} = \mathbf{F}$. Let ($\theta_x$; $\theta_y$; $\theta_z$) denote angular displacements around the (X; Y; Z)-axes, respectively (Fig2.). Then $\boldsymbol{\omega}$= ($\omega_x$; $\omega_y$; $\omega_z$)$^T$=($\dot{\theta}_x$; $\dot{\theta}_y$; $\dot{\theta}_z$)$^T$ is the vector of angular velocity. In this configuration the principal moments of inertia are represented in the matrix

$$\hat{I} = \begin{pmatrix} I_1 & 0 & 0 \\ 0 & I_2 & 0 \\ 0 & 0 & I_3 \end{pmatrix}, \quad I_1 = I_2 = m_{ls}L_c^2 + m_{sc}(L-L_c)^2 = I,$$

(the cilindrical symmetry is taken into account).

Since the spacecraft is assumed to be rigid, its motion can be described by Newton's second law $m\ddot{\mathbf{X}} = \mathbf{F}$ and Euler's equation $I\dot{\boldsymbol{\omega}} + \boldsymbol{\omega} \times I\boldsymbol{\omega} = \mathbf{M}$.

The pondero-motive force $\mathbf{F}$ represents the radiation pressure produced by a laser beam with intensity vector on a mirror reflective sail of area S and is given by

$$\mathbf{F} = 2\int_S (\mathbf{w}(\rho)\mathbf{n}(x,y))\mathbf{n}(x,y)dS \quad , \tag{1.1}$$

where coefficient 2 accounts for the contribution of reflected beam, $\mathbf{n}(x,y)$ is the unit vector normal to the sail surface at the point $(x,y)$ (counted relative vertex point in the frame of reference of the sail), $w=(0, 0, w(\rho))$. For spherical shape of sail with $R$ being its radius of curvature $(\mathbf{w}(\rho)\mathbf{n}(x,y)) = w(\rho)\dfrac{\sqrt{R^2 - x^2 - y^2}}{R}$ (here we assume the angle of inclination of sail axes to be small, thus, consistent with linear approximation). In principal (undisturbed) position of sail, when axes of symmetry of the sail and beam coincide, $\mathbf{F}_0$ is in direction of laser beam.

Now lets introduce the torque $\mathbf{M}$ caused by the beam on the sail

$$\mathbf{M} = 2\int_S w(\rho)\frac{\sqrt{R^2 - x^2 - y^2}}{R}[\mathbf{L}_c(\mathbf{x}) \times \mathbf{n}(x)]dS \quad , \tag{1.2}$$

where $\mathbf{L}_c(\mathbf{x})$ is the vector from the center of mass of nanocraft to the point $\mathbf{x}$ on the sale. The net torque $\mathbf{M}_0$ in principal position of sail vanishes.

In a linear stability analysis we will deal with small deviations of the center of mass X and Y and the small inclination of the spacecraft's principal axis of symmetry from the Z-axis, which we will associate with a rotation angles $\theta_x$ and $\theta_y$ about the center of mass. Below we consider linear approximation of nanocraft position stability inside laser illuminated column depending on the shape of the sail and the intensity profile of laser beam. There are two contributing factors to cause imbalance in torque and force when the sail position and orientation are perturbed: first, the net shift in lateral direction (denoted as Δ), which takes a sail out of original central position inside the illuminated column, thus, leading to imbalanced torque; and the second, actual turn of sail, its angular inclination leading to imbalance in lateral force and torque. We will consider them separately before summing up contributions of both in final expressions for the torque and force.

In the perturbation shown in Fig. 4 the right edge of sail, displaced by distance Δ in positive X direction, goes outside the laser illuminated area. Perturbation in Y direction is decoupled from the one in X in the absence of spinning around axis. Thus, the stability analysis is reduced to the case of 2 degrees of freedom (motion along X and rotation around Y axis, or similarly to motion along Y and rotation around X axis). We will take the case for X and $\theta_y$. The distance Δ includes contributions from spacecraft shifting in the X direction and turning clockwise around the Y-axis. In linear approximation we have $\Delta = X + L_r\theta_y$, where $L_r = \sqrt{(\sqrt{R^2 - a^2} + L_c - R)^2 + a^2}$ is the distance between center of mass and rim of the sail. The shift of the sail away from the axis of the beam leads to imbalance in the torque in comparison to original alignment, and similarly to imbalance in the force in X direction.

Now we have to specify $\mathbf{L_c}$ - the vector from center of mass of nanocraft to the point (x,y) on sail surface: $\mathbf{L}_c(x, y) = \{x, y, L_c - R + \sqrt{R^2 - x^2 - y^2}\}$ where $L_c$ the distance from center of mass to the vertex of the sail. With the use of definition of $\mathbf{n} \cong \{\frac{x}{R}, \frac{y}{R}, \frac{\sqrt{R^2 - x^2 - y^2}}{R}\}$ we get the expression for the y-component of local torque $dM_y$ on element of the sail surface $dS$ (see Fig. 5) is

$$dM_y = 2w(\rho) \frac{\sqrt{R^2 - x^2 - y^2}}{R} (\frac{L_c}{R} - 1) x dS . \quad (1.3)$$

For chosen model of laser illumination (constant intensity at $\rho < a$) $w(\rho) = w_0$ everywhere over the sail surface except the right rim of it taken out of illumination. The square area of the $dy$ element of this rim is $dS = \Delta \cdot dy \frac{R}{\sqrt{R^2 - x^2 - y^2}}$. In a new position of sail it is the torque acting on the left rim, which is not counterbalanced on the opposite (right) side. Now after substituting it in expression (1.3) we get for the net torque acting on perturbed nanocraft

$$M_y = 2w_0(\frac{L_c}{R} - 1)\int_S xdy\Delta = -2w_0\Delta(\frac{L_c}{R} - 1)\int_{-a}^{a} \sqrt{a^2 - y^2}\, dy = -\frac{1}{2}F_{rad}(\frac{L_c}{R} - 1)(X + L_r\theta_y), \quad (1.4)$$

where $F_{rad} = 2w_0\pi a^2$ and we took into account that $x = -\sqrt{a^2 - y^2}$ (for the left side of sail). Note that the sign of torque corresponds to restoring action if $L_c > R$.

Similarly, the force in the X direction due to the shift of sail in this direction can be calculated as:

$$F_x == -\frac{1}{2}F_{rad}\frac{X + L_r\theta_y}{R}. \quad (1.5)$$

Additional force is associated with the turn on angle $\theta_y$ which does not take the sail out of illumination, but simply creates x component of force due to inclination of sail $F_{rad}\theta_y$.

In general case when $a$ is comparable to $R$, there are also contributions to torque and force caused by the turn (per se) of sail around Y-axis on angle $\theta_y$. In order to simplify the arguments about stability we are restricting following calculations by assuming $a^2$ is fined but much less then $R^2$. In this approximation we can consider $L_r = L_c$.

Euler's equation and Newton's second law can be written in the form:

$$I\ddot{\theta}_y = -\frac{1}{2}F_{rad}(\frac{L_c}{R} - 1)(X + L_c\theta_y), \quad (1.6)$$

$$m\ddot{X} = F_{rad}\theta_y - \frac{1}{2}F_{rad}\frac{1}{R}(X + L_c\theta_y). \tag{1.7}$$

Equations for $\theta_x$ and $Y$ are similar to (1.6), (1.7). The coupling between $\theta_x$ and $\theta_y$ enters only via spinning around Z axis. We will consider it later.

In linear stability analysis we look for solutions of (1.6), (1.7) of the form

$$\theta_y = \tilde{\theta}_y e^{i\nu t}, X = \tilde{X}e^{i\nu t}. \tag{1.8}$$

The solvability condition (dispersion equation) for the case $m_{ls}=m_{sc}$ and $a \ll R$ (almost flat sail) is reduced to $I = \frac{mL^2}{4}$, $I_3 = m\frac{a^2}{2} \ll I$,

$$\nu^4 + \frac{F_{rad}}{2mL_c}(1 - \frac{2L_c}{R})\nu^2 - \frac{F_{rad}^2}{2m^2L_c^2}(1 - \frac{L_c}{R}) = 0, \tag{1.9}$$

$$\nu^2 = \frac{F}{2mL_c}, \frac{(L_c - R)F}{RmL_c}.$$

If $L_c > R$ then the nanocraft orientation is stable. Criteria of stability is exactly the same even without assumption about almost flat sail, though the expressions for eigenvalues are more cumbersome. Thus, stability is achieved when the distance from the lightsail center to the center of mass is larger than the radius of curvature.

If the sail rotates around its axis Z with angular frequency $\omega$, then we have coupling between motions and rotation in previously two decoupled systems

$$I\ddot{\theta}_y + (I_3 - I)\dot{\theta}_x\omega - \frac{1}{2}F_{rad}(X + L_c\theta_y) + \frac{1}{2}F_{rad}\frac{L_c}{R}(X + L_c\theta_y) = 0, \tag{1.10}$$

$$I_2\ddot{\theta}_x + (I - I_3)\dot{\theta}_y\omega - \frac{1}{2}F_{rad}(Y + L_c\theta_x) + \frac{1}{2}F_{rad}\frac{L_c}{R}(Y + L_c\theta_x) = 0, \tag{1.11}$$

$$m\ddot{X} = F_{rad}\theta_y - \frac{1}{2}F_{rad}\frac{1}{R}(X + L_c\theta_y), \tag{1.12}$$

$$m\ddot{Y} = F_{rad}\theta_x - \frac{1}{2}F_{rad}\frac{1}{R}(Y + L_c\theta_x). \tag{1.13}$$

Here we assume $a \ll L$ and $m_{ls}=m_{sc}$, so $I = \frac{mL^2}{4}$, $I_3 = m\frac{a^2}{2} \ll I$. The solvability condition (dispersion equation) stemming from (1.10-1.13) is

$$\Omega^4 + \frac{2F_{rad}}{mL}(1 - \frac{L}{R} - \frac{Lm\omega^2}{2F_{rad}})\Omega^3 - 3\left(\frac{F_{rad}}{mL}\right)^2\left(1 - \frac{1}{3}\left(\frac{L}{R}\right)^2 - \frac{L^2m\omega^2}{3RF_{rad}}\right)\Omega^2 -$$
$$- 4\left(\frac{F_{rad}}{mL}\right)^3\left(1 - \frac{3}{2}\frac{L}{R} + \frac{1}{2}\left(\frac{L}{R}\right)^2 + \frac{L^3m\omega^2}{16R^2F_{rad}}\right)\Omega + 4\left(\frac{F_{rad}}{mL}\right)^4\left(1 - \frac{L}{R} + \frac{1}{4}\left(\frac{L}{R}\right)^2\right) = 0$$
(1.14)

where $v^2 = \Omega$.

Rotation (spinning) of the craft around Z axis can bring the stability even if $L<2R$, but at the cost which might require a too high frequency of spinning.

## 2. Conical shape of the sail at $a \ll L$ for the same flat laser beam radial profile with sharp boundary.

The same type of linear stability analysis as above can be done for conical sail. It is very simple at the assumption of an almost flat cone (angle $\alpha \ll 1$, see Fig. 6). The perturbation of sail position and orientation takes the edge of sail out of the beam column. The slope of the conical sail is defined by the angle $\alpha$.

The line perpendicular to surface of sail at the edge crosses Z-axis at the distance $R = \frac{a}{\sin\alpha}$. In case of spherical sail that coincides with the curvature radius if $\alpha$ would be the slope angle of spherical sail at the edge. Here by using the analogy with spherical cone we can conclude that stability is achieved If $L_c > 2a/\sin\alpha$.

## 3. Flat sail (particular case).

The assumptions in choosing the model are: 1) configuration of nanocraft is treated as solid body (applicability of Euler equations); 2) flat round sail; 3) laser beam is uniform with sharp boundary; 4) mirror reflection of laser beam from surface of the Lightsail.

$$I\ddot{\theta}_y - \frac{1}{2}F_{rad}(X + L_c\theta_y) = 0,$$ (3.1)

$$m\ddot{X} = F_{rad}\theta_y.$$ (3.2)

Equations for $\theta_x$ and $Y$ are similar to (3.1), (3.2). The solvability condition (dispersion equation) for case $m_{ls}=m_{sc}$ is

$$v^4 + \frac{F_{rad}}{2mL_c}v^2 - \frac{F_{rad}^2}{2m^2L_c^2} = 0,$$ (3.3)

$$v^2 = \frac{F}{2mL_c}, -\frac{F}{mL_c}.$$

The nanocraft is unstable in every case. It simply corresponds to the case of infinite radius of curvature of the sale. Hence, there is no way to extend center of mass beyond it.

## 4. Gaussian beam shape and spherical sail.

Now we will consider the radial shape of laser beam as Gaussian (to emulate diffraction controlled shape of beam). Thus, $w(\rho) = w_0 e^{-\frac{\rho^2}{r^2}}$, where $\rho$ is the distance from the beam axis Z and $r$ indicates the spread of the beam intensity. In its equilibrium position the sail is occupying central position in the beam, so that maximum of beam intensity falls on the center of sail. Perturbation of its position in a form of shift in X direction takes the sail out of symmetry in illumination and hence causes the imbalanced torque and lateral force. Here we will limit ourselves with assumption of almost flat sail like in previous chapter.

Lets take the displacement of sail as $\Delta = X + L_r \theta_y$ and calculate the imbalance in torque around X axis and a force along X. When shifted by distance $\Delta$ the sail is exposed to laser illumination radial profile $w_0 \exp\left[-\frac{(x-\Delta)^2 + y^2}{r^2}\right]$. In linear stability analysis we can expand it as $w_0 \exp\left[-\frac{x^2 + y^2}{r^2}\right].(1 - 2x\Delta)$. Then for the local torque over the element of sail surface we get $dM_y = -2dS w_0 \exp\left[-\frac{x^2 + y^2}{r^2}\right] x\Delta [\mathbf{L}_c \times \mathbf{n}]$. Using familiar expressions for vectors $\mathbf{L_c}$ and $\mathbf{n}$ and integrating over the surface of sail we find the torque $M_y$ related to shift along X-axis:

$$M_y = -w_0 \frac{\Delta}{r^2} \int_S x^2 \exp\{-\frac{x^2 + y^2}{r^2}\} dxdy = -w_0 (\frac{L_c}{R} - 1) \frac{\Delta}{r^2} \int_0^a \int_0^{2\pi} \rho^2 \cos^2\varphi\, e^{-\frac{\rho^2}{r^2}} \rho d\rho d\varphi =$$

$$= -\frac{1}{2} w_0 \pi a^2 \frac{r^2}{a^2} (\frac{L_c}{R} - 1)(X + L_c \theta_y) \int_0^{\frac{a^2}{r^2}} t e^{-t} dt .$$

Now let us write

$$F_{rad}^* = w_0 \pi a^2 \frac{r^2}{a^2} \int_0^{\frac{a^2}{R^2}} t e^{-t} dt .$$

We can get expression for tangential restoring force $f_x$:

$$f_x = -\frac{1}{2} F_{rad}^* \frac{1}{R}(L_c \theta_y + X), \quad F_{rad}^{**} = F_{rad}^* \frac{\int_0^{\frac{a^2}{R^2}} e^{-t} dt}{\int_0^{\frac{a^2}{R^2}} t e^{-t} dt}.$$

Euler's equation and Newton's second law can be written in the form:

$$I\ddot{\theta}_y - F_{rad}^*(X + L_c\theta_y) + F_{rad}^* \frac{L_c}{R}(X + L_c\theta_y) = 0, \qquad (4.1)$$

$$m\ddot{X} = F_{rad}^{**}\theta_y - \frac{1}{2}F_{rad}^* \frac{1}{R}(X + L_c\theta_y). \qquad (4.2)$$

In linear stability analysis we look for solutions of (4.1;4.2) as in (1.8). The solvability condition (dispersion equation) is

$$v^4 + \frac{F_{rad}^*}{mL}(1 - \frac{L}{R})v^2 - \frac{2F_{rad}^* F_{rad}^{**}}{m^2 L^2}(1 - \frac{L}{2R}) = 0. \qquad (4.3)$$

If $L > 2R$ then the nanocraft orientation is stable with no need for spinning.

## 5. Numerical calculations.

The Lightsail is assumed to be a part of a sphere with mass m=0.5 g and radius $a$. StarChips mass is the same m=0.5 g, $L$ is the distance between the sail and chip, $a$ =200 cm is the shortest distance between the sail edge and the axis connecting the sail and the chip. We assume that the sail is slightly curved, so $r \approx a$. $R$ is the radius of curvature of the sail and $a<<R$.

We model the intensity $w$ of the laser beam as constant inside the cylinder with radius $R$ (same as of lightsail) and which acts on the full disk with the force $w \times \pi R^2 = F_{rad}$. This force have to accelerate disk up to speed $v_{max} = 6 \times 10^9$ cm/s during time interval $\Delta t$ =120 s. Therefore $F_{rad} = 5 \times 10^7$ g×cm/s$^2$.

Solving Eq.(1.6 - 1.7) together with similar equations for $\theta_x$ and Y, numerically we obtain results confirming our analytical calculations. In Fig. 7 the position of the nanocraft is presented during beam-riding simulation in the case of spherical shape of Lightsail, uniform laser beam with sharp boundary, L=2000 and R=1000. For initial perturbations of coordinates $x$ and $y$ and angles = 0.001 m amplitude of oscillations of sail position ≈ 1 cm. For initial perturbations of coordinates and angles = 0.01 m amplitude of oscillations of sail position ≈ 10 cm. With further increase of the initial perturbations of coordinates the amplitude of oscillations are quite large. The presence of ω does not affect the amplitude of oscillations.

III DISCUSSION

Our results differ from [11]. Note that in contrast to that paper, we analyzed problem of stability for configuration of nanocraft where sail and chip are separated in space.

For cases of spherical and conical sail shapes, in contrast to the model in [11], we present detailed stability analysis on dependence of the crucial parameters of the model. The stability is achieved when the distance from the Lightsail center to the center of mass is larger than the radius of curvature. This fact has simple physical meaning.

In Fig. 8 the sail and part of the suspension line connecting the center of mass and the sail is shown by red color. In the case $L_c > R$ the torque returns the sail to its initial position, if $L_c < R$ the torque pushes the sail out of the beam.

**Acknowledgements**

The authors express their gratitude to Dr. Roald Sagdeev for introducing us to this fascinating subject, as well as both to him and Dr. Arif Pashayev for very useful scientific discussions. Moreover we all enjoyed the support and hospitality of the National Aviation Academy of Azerbaijan.

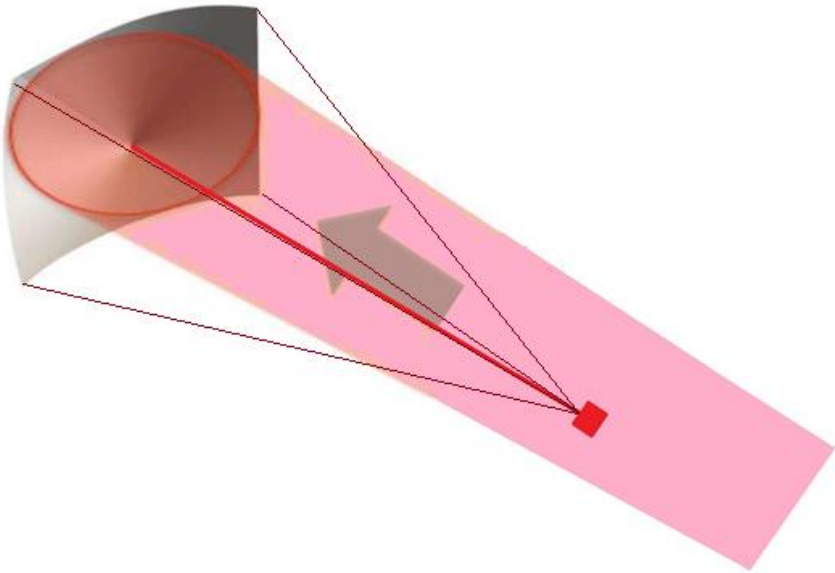

Fig. 1. Schematic illustration of the Nanocraft.

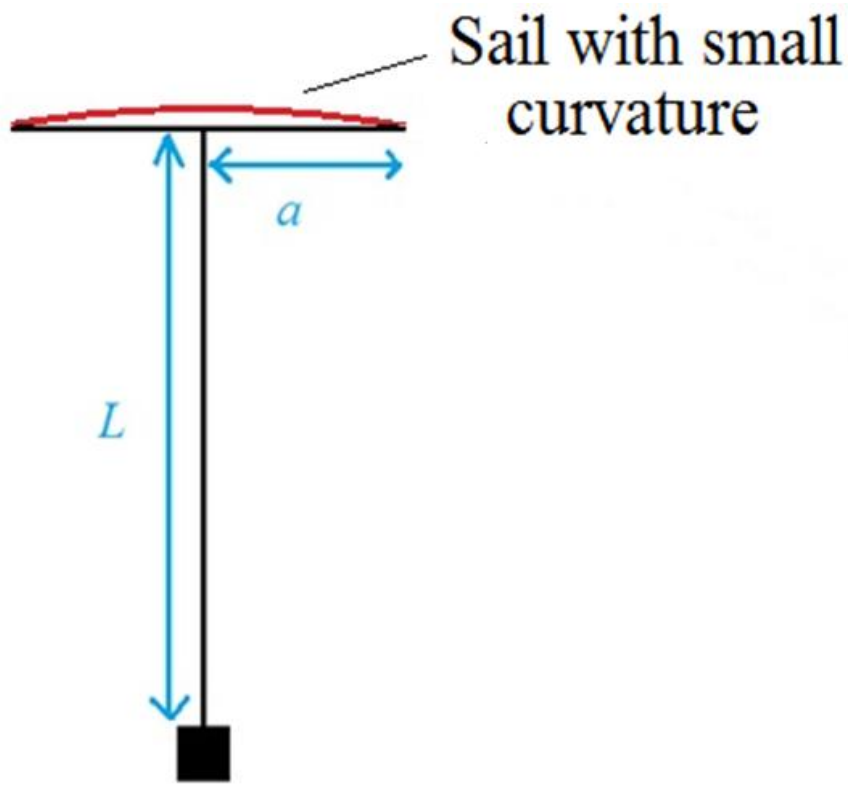

Fig. 2. Scheme of nanocraft with sail as part of sphere.

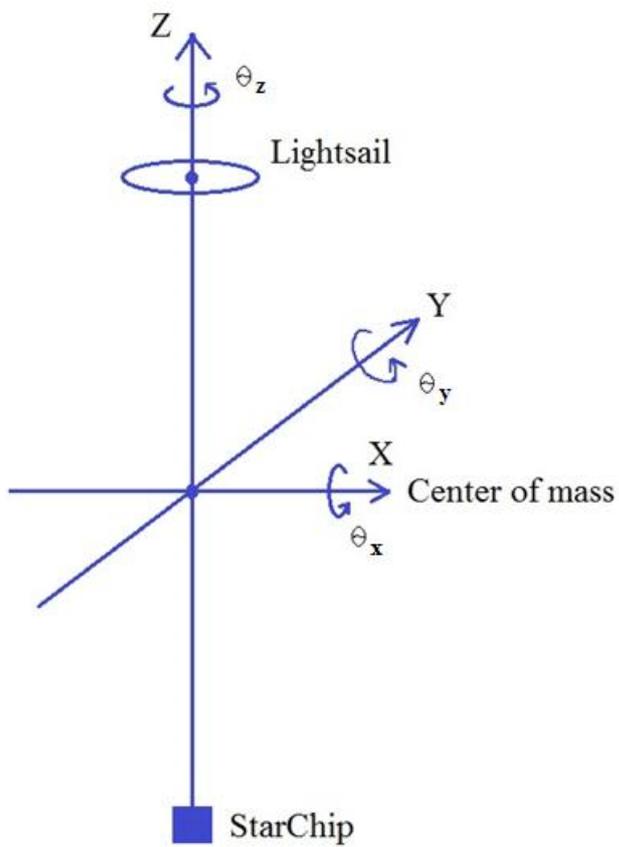

Fig. 3. Introduction of principal axes, Eulerian angles and center of mass of nanocraft.

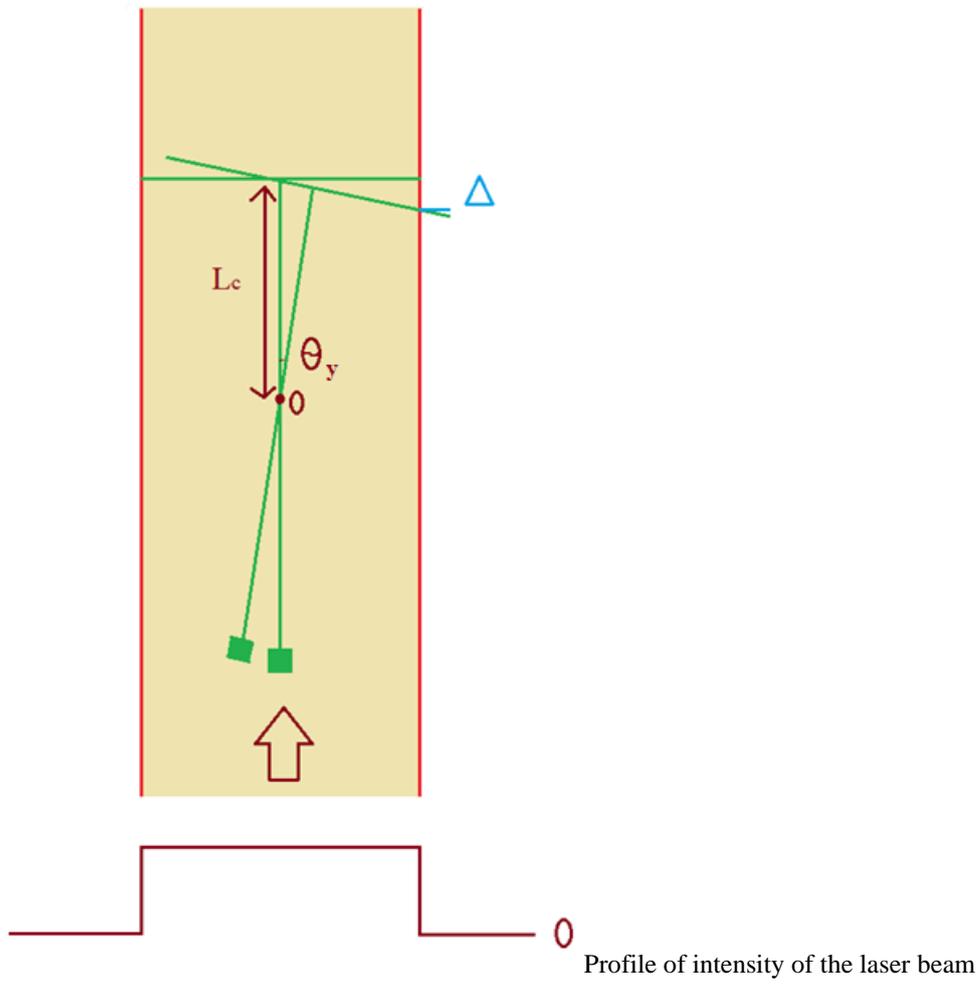

Profile of intensity of the laser beam

Fig.4. Origin of the torque as result of lateral displacement.

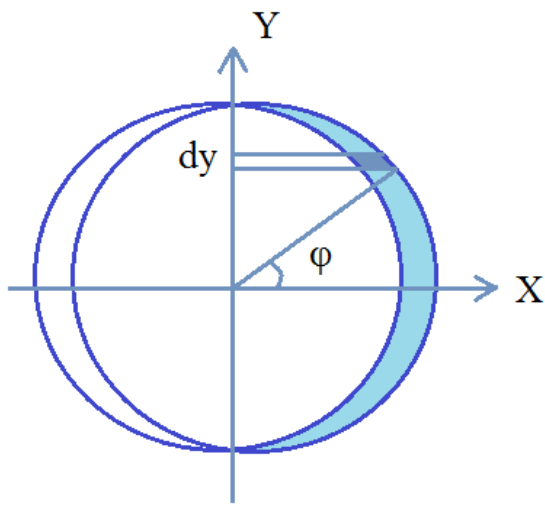

Fig. 5. Scheme for calculations of torque disbalance.

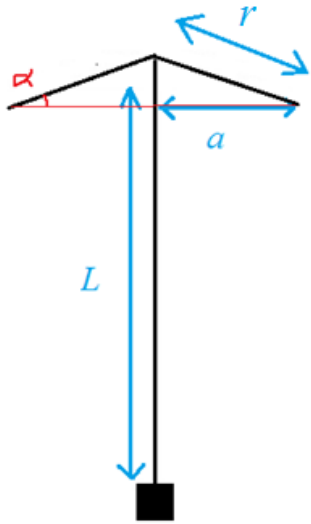

Fig. 6. Conical shape of the sail.

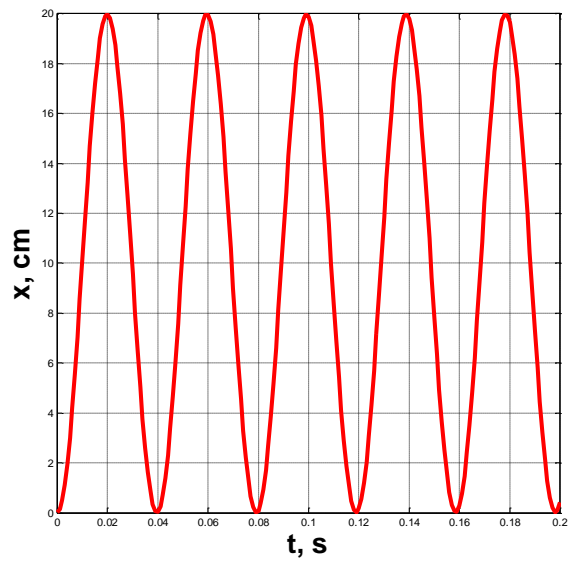

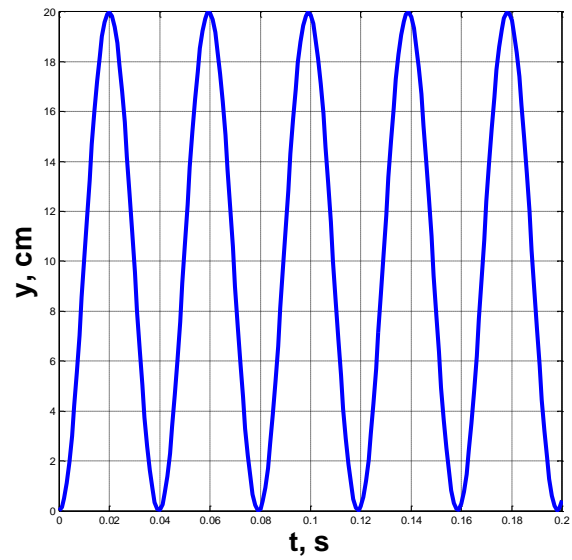

Fig. 7. Sail position during beam-riding simulation.

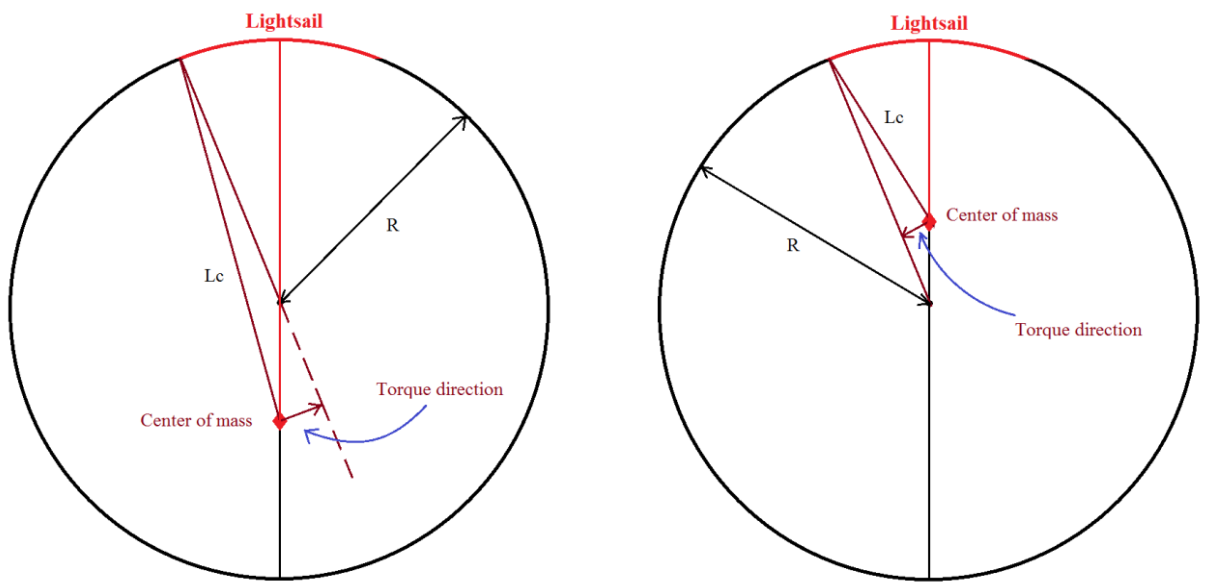

Fig. 8. Scheme of torque directions in case of $L_c > R$ (left panel) and $L_c < R$ (right panel).

15